\documentclass[lettersize,journal]{IEEEtran}

\usepackage{cite}
\usepackage{amsmath,amssymb,amsfonts}
\usepackage{algorithmic}
\usepackage{graphicx}
\usepackage{textcomp}
\usepackage{graphics,color, float, epsf, caption, subcaption}
\usepackage{tabularx}
\usepackage[labelsep=period]{caption}
\renewcommand{\thetable}{\Roman{table}}
\usepackage{tabularray}
\usepackage{cuted}
\usepackage{stfloats}
\usepackage{hyperref}

\def\BibTeX{{\rm B\kern-.05em{\sc i\kern-.025em b}\kern-.08em
    T\kern-.1667em\lower.7ex\hbox{E}\kern-.125emX}}
\begin{document}

\title{Dynamic Electric Vehicle Charging Pricing for Load Balancing in Power Distribution Networks based on Collaborative DDPG Agents}

\author{Leloko J. Lepolesa, Kayode E. Adetunji, Khmaies Ouahada, Zhenqing Liu and Ling Cheng, ~\IEEEmembership{Senior Member,~IEEE,}}


\markboth{IEEE TRANSACTIONS ON SMART GRID,~Vol.~..., No.~..., ~2025}%
{Shell \MakeLowercase{\textit{et al.}}: A Sample Article Using IEEEtran.cls for IEEE Journals}


\maketitle

\begin{abstract}

The transition from the Internal Combustion Engine Vehicles (ICEVs) to the Electric Vehicles (EVs) is globally recommended to combat the unfavourable environmental conditions caused by reliance on fossil fuels.
However, it has been established that the charging of EVs can destabilize the grid when they penetrate the market in large numbers, especially in grids that were not initially built to handle the load from the charging of EVs. In this work, we present a dynamic EV charging pricing strategy that fulfills the following three objectives: distribution network-level load peak-shaving, valley-filling, and load balancing across distribution networks.
Based on historical environmental variables such as temperature, humidity, wind speed, EV charging prices and distribution of vehicles in different areas in different times of the day, we first forecast the distribution network load demand, and then use deep reinforcement learning approach to set the optimal dynamic EV charging price.
While most research seeks to achieve load peak-shaving and valley-filling to stabilize the grid, our work goes further into exploring the load-balancing between the distribution networks in the close vicinity to each other. We compare the performance of Deep Deterministic Policy Gradient (DDPG), Soft Actor-Critic (SAC) and Proximal Policy Optimization (PPO) algorithms for this purpose. The best algorithm is used for dymamic EV pricing. Simulation results show an improved utilization of the grid at the distribution network level, leading to the optimal usage of the grid on a larger scale.

\end{abstract}


\section{Introduction}
\label{sec:introduction}

\IEEEPARstart{T}{he} Electric Vehicles (EVs) are rapidly penetrating the market. The main reason for adopting the EVs in place of Internal‐Combustion Engine (ICE) powered vehicles is to reduce CO2 emissions from end‐use sectors, of which the transport sector is reported to account for one third \cite{iea_co2}. In \cite{iea_co2}, it is further reported that the sales of EVs grew by 55\% in 2022, attaining a record high of more than 10 million.
In as much as the decarbonisation of the transport sector brings environmental benefits, it is worth considering that many distribution networks were not originally designed to support EV load. As a result, integrating the charging of EVs into these networks introduces an additional load, potentially leading to grid instability \cite{tariang2023survey}. 

Researchers in \cite{strezoski2024enabling} and \cite{nutkani2024impact} mention the challenges of mass EV penetration into the grid, that include increased peak load demand, increased voltage violations, increased system imbalance due to single phase chargers, increased power losses due to increased rate of consumption, and overloading of transformers and distribution feeders, which in turn will violate their thermal limits, thereby reducing their lifespan. 
In \cite{li2024impact}, EV charging data was gathered from the utilities, public-charging service providers, and the data logger records in California state. It was combined with travel behaviour data to find the impact of EVs on the distribution grids. It was predicted that 50\% of feeders must be upgraded by 2035, and 67\% by 2045.

In another case study made by Abiassaf and Arkadan \cite{abiassaf2024impact}, the authors investigated the impact of EV charging on the distribution grid in Colorado Springs. By utilizing the household data, vehicle technology, passenger travel, and fuel types, they used the Transportation Energy and Mobility Pathway Options (TEMPO) model to predict the impact of EV charging on the distribution network in the next 30 years. They report that by 2050, the impact will be severe.

To mitigate this, coordinated EV charging strategies are essential for minimizing their adverse effects on the grid and enhancing its overall efficiency \cite{aghajan2022charging}.
To ensure the optimal grid operation and the satisfaction of EV users, careful considerations have to be made on when and where the EVs are charged on the grid.
Researchers in  \cite{qu2024physics, kuang2024physics, kuang2024unravelling} conducted studies that show the EV load demand behaviour when the prices are varied in different areas. Better road networks and the usage of smartphones facilitate the change in the EV charging demand in the neighbouring areas when changes in EV charging price are made in one area \cite{kuang2024physics}. Therefore it is apparent that when the state of charge reaches the threshold required for an EV to be charged, factors that influence the area of charging for the user include the distance to the Electric Vehicle Charging Station (EVCS) and the EV charging price at that station.

The existing research on scheduling the charging of electric vehicles for grid  stability focuses on load peak-shaving and valley-filling. 
Upon studying load profiles at the level of distribution networks, it is observed that different distribution networks have different patterns of power consumption.
While conventional Time-of-Use (ToU) pricing is intended to improve utilization of the grid, implementing a mechanism to optimize grid usage at the level of the distribution networks can further enhance grid efficiency and stability. If the conventional ToU pricing structure is applied to the charging of EVs, a new electricity demand peak is created when the charging price is lowest in the residential areas, as studied in \cite{kim2019insights}. In the commercial areas, authors in \cite{gilleran2021impact} found that the EV load peaks between 12:00 and 18:00 when the conventional load is also at peak, which can be a major issue if the capacity of the distribution network is reached.

The integration of EVs into the power grid has the potential to cause problems in the grid if not well managed, hence the scheduling of charging and discharging of EVs has attracted many researchers globally.
Researchers in \cite{srividhya2024optimizing} propose a hub synchronization method that uses dynamic K-means clustering for optimized EV charging. The EV charging data is analysed continuously to recognize the patterns in user conduct, and then predicts the future demand based on the historical data and reallocates EVs to have optimal distribution at the charging locations. However, the proposed method does not show how the users are convinced to charge at the different locations.


Authors in \cite{chakraborty2024planning} use Multi-Objective Particle Swarm Optimization (MOPSO) to design an optimal placement of EV Charging Infrastructure (EVCI) in a manner that minimizes the power loss and voltage deviation in a distribution network. They further propose a distribution network-level dynamic pricing strategy that can be predicted using Autoregressive Integrated Moving Average (ARIMA), to improve the grid stability, efficiency and revenue.
In \cite{kazemtarghi2024dynamic}, 
authors formulate the CS attraction function to the EV based on Reilly's law of retail gravity, and suggest charging prices for each CS based on scenario-based stochastic optimization. While the method improves the CSO revenue and relieves the congestion of idividual charging stations in comparison to the fixed pricing strategy, the distribution networks congestion is not addressed.

Das and Kayal \cite{das2024advantageous} forecasts a day-ahead PV energy and propose a surface estimation (SE) charging and discharging scheduling algorithm to minimize the grid load variation, which outperforms water filling (WF) algorithm interms of computation time.
Authors in \cite{nath2024short} presented short-term EV charging load forecasting using Transfer Learning (TL) and Model-Agnostic Meta-Learning (MAML), for small datasets. Three datasets are used in this research work. The larger dataset is used to pre-train the TL and MAML models, which are then fine-tuned with the smaller datasets on 10-day and 20-day subsets to predict the hourly load for the next 10 days and 20 days, respectively. They compared the performance of TL and MAML models with the custom Long Short-Term Memory (LSTM) deep learning model, where the TL and MAML models showed their superiority over the LSTM model.

Authors in \cite{kuang2024unravelling} studied the impact of electricity price on EVs charging behaviour. They analysed the month-long dataset of public EV charging piles in Shenzhen city, China. With varying electricity prices in CBD, the spillover of 89.48\% was observed when the charging price went high, with an estimated spillover radius of 3.45km. In \cite{lee2018analysis}, a game theory approach was used to model the price competition between multiple small-sized EVCSs and one large-sized EVCS with renewable power generators. Through simulation validation, they established that the capacity of EVCS, charging price, and distance determine the charging behaviour of EVs, and that the EVCS have a better payoff than when the power from renewable energy generators is sold to the grid.

Saxena and Gao \cite{saxena2024power} consider the random fleet of EVs at the EVCS, and use the Distribution Static Synchronous Compensator (D-STATCOM) to balance the load at the EVCS.Through careful planning of charging gun occupancy cycles, transients, current harmonic distortions, and current and voltage imbalances are improved.
The presented methods show satisfactory performance in optimal integration of the EVs in the power grids; however, the research on distributing the EV charging load demand among the neighbouring distribution networks is less explored. The new approach introduced in this work considers the distribution network-level power consumption for EV charging pricing strategy. Our method relies on the distribution network load power consumption patterns, contrary to the methods in the literature, which consider multiple inputs that include the attributes of charging stations, such as the number of charging slots, EVCS charging speed, and attributes of electric vehicles, such as the State of Charge (SoC).  
Li et al. \cite{li2024toward} describe the different types of EVs and their charging modes. These also have to be considered when individual vehicles' charging behaviours are taken into account as inputs to optimal charging scheduling methods.

In the literature, the EV charging pricing strategies proposed significantly cover peak-shaving and valley-filling in the power consumption pattern. Our EV charging framework further balances the utilization between distribution networks that are in close proximity. In our previous work \cite{lepolesa2024optimal}, we proposed a methodology that utilizes the discrete time of use (TOU) prices to set the dynamically changing EV charging prices. This work proposes continuous price increments to existing conventional TOU prices, based on the load pattern in a distribution network, and the relative difference in utilization with the neighboring distribution networks.

On a sample of 237 early EV adopters in Germany, Will and Schuller \cite{will2016understanding} found that leaving with a full battery is important to users, which is what our method offers, in addition to the charging price advantages.
The contributions are as follows:

\begin{itemize}
    \item We introduce a novel method to optimally use the power grid at the distribution network level by leveraging EVs as mobile loads that can be charged wherever it is convenient for the EV user. While most research seeks to achieve load peak-shaving and valley-filling to stabilize the grid, our work goes further into redirecting EV charging demand to nearby distribution networks based on charging price incentives, thereby achieving the load-balancing between the distribution networks in close vicinity, and further optimizing the grid usage on a larger scale.
    
    \item Contrary to the methodologies in the literature that consider multiple inputs such as SoC of individual electric vehicles, number of charging slots and the associated charging rates in charging stations, etc, we design the first electric vehicle charging pricing scheme that takes only the distribution network load and collective distribution of EVs at the given time as the input, and provides the EV owner freedom to charge an EV as long as they wish, to the SoC level required by them.
    
    \item By utilizing real-world distribution network-level data and the distribution of vehicles in different times of the day, we demonstrate that both the conventional load and EV load data can be accurately forecasted, thereby forming the basis for the informed day-ahead dynamic EV charging pricing that is suitable for optimizing the load demand patterns.
    
    \item We design a collaborative deep reinforcement learning (DRL) reward function that allows the dynamic EV charging price-setting agents to collaboratively control the EV load in each distribution network.    
\end{itemize}

The remainder of this paper is as follows:  Section \ref{sec: background} presents the background information on the following algorithms that play a critical role in the methodology presented in this work: Extreme Gradient Boosting (XGBoost), linear regression, and Deep Deterministic Policy Gradient (DDPG) deep reinforcement learning. Section \ref{sec:method} outlines the proposed method that combines the forecasting of the distribution network load and dynamic EV charging pricing to enhance EV charging in distribution networks. In Section \ref{sec:discussion}, we present and discuss the results that demonstrate the effectiveness of the proposed scheme. We finally conclude the paper and state the future research directions in Section \ref{sec:conclusion}.


\section{Background}
\label{sec: background}

Different techniques were used for distribution network load forecasting and dynamic pricing. For the forecasting of conventional distribution network load, the performances of feed-forward neural networks, XGBoost, and polynomial regression are compared. The derivation of the charging load of EVs is in the form of polynomial regression. For the price-setting DRL agents, we compare the performances of Deep Deterministic Policy Gradient (DDPG), Soft Actor-Critic (SAC) and Proximal Policy Optimization (PPO) algorithms. In this section, we present a brief background of the techniques that yield the best results, which are XGBoost, polynomial regression, and the DDPG reinforcement learning algorithm.

\subsection{Extreme Gradient Boosting (XGBoost)}

Extreme gradient boosting (XGBoost) is a scalable machine learning system for tree boosting, introduced by Tianqi Chen and Carlos Guestrin \cite{chen2016xgboost}. It has been used successfully since 2015 after its inception, outperforming other classification and regression models in competition platforms like Kaggle. 
It uses a gradient tree boosting mechanism to make predictions. It optimizes an objective function $f$ given by \eqref{eq: supervised f}.
\begin{equation} \label{eq: supervised f}
    L^{(k)} = \sum_{i=1}^{n}l(y_i,\hat{y}_i^{(k-1)}+f_k(\mathbf{x}_i)+\Omega {(f_k)},
\end{equation}
where $l$ is a differentiable convex loss function that measures the difference between the actual value $y_i$ and the sum of predictions from previous trees $\hat{y}_i^{(k-1)}$ and the new tree predictions $f_k(\mathbf{x}_i)$. $\Omega$ is the regularization term that penalizes the complexity of the model, given by \eqref{eq: omega}:
\begin{equation} \label{eq: omega}
    \Omega {(f)} = \gamma T + \frac{1}{2}\lambda\sum\alpha_j^2,
\end{equation}
where $T$ is the number of leaves, $\alpha_j$ is the weight of a $j^{th}$ leaf, and $\gamma$ and $\lambda$ are regularization parameters.
Second-order Taylor expansion is used to approximate $L^{(k)}$. The split is determined by Gain, given by \eqref{eq: gain}:
\begin{equation} \label{eq: gain}
    \text{Gain} = \frac{1}{2}\left( \frac{G_L^2}{H_L+\lambda }+ \frac{G_R^2}{H_R+\lambda} - \frac{(G_L + G_R)^2}{H_L+H_R+\lambda }\right)-\gamma,
\end{equation}
where $G_L$ and $G_R$ are sums of gradients for the left and right child nodes, and $H_L$ and $H_R$ are sums of the Hessians for the left and right child nodes, respectively.

One major difference between XGBoost and other supervised algorithms is that it focuses on out-of-core computation and cache-aware learning, instead of the algorithmic aspect of parallelization. This allows for the ability to handle large scale problems using the limited computing resources.
In this work, we use XGBoost for the future prediction of power consumption by the conventional load in the distribution network, using environmental variables as the predictors.

\subsection{Polynomial regression}

Polynomial regression is used to model non-linear relationships between predictors and response variables by fitting a polynomial equation to the data. Polynomial regression is in the form $y = f(\mathbf{x}, \mathbf{\beta}) + \epsilon$, where $\mathbf{x}$ is the predictors, $\mathbf{\beta}$ is the parameters to be estimated, and $y$ is the response variable \cite{chen2014polynomial}. For a single predictor $x$, polynomial regression is in the form $y = \beta_0 + \beta_1x +\dots+\beta_kx^k+\epsilon$, where $k$ is the order of the model. To avoid overfitting, the order of the model is kept at the lowest value that produces acceptable results.

We use polynomial regression to make future predictions of power consumption in a distribution network by electric vehicles.

\subsection{Deep Reinforcement Learning (DRL) - Deep Deterministic Policy Gradient (DDPG) algorithm}

Deep Reinforcement Learning (DRL) methods incorporate Deep Neural Networks (DNNs) to approximate any of the following: value function $V(\mathbf{s};\theta)$, Q-function $Q(\mathbf{s},\mathbf{a};\theta)$, policy $\pi(\mathbf{a}|\mathbf{s};\theta)$, state transition function and the reward function $R$ \cite{li2017deep,wang2022deep}, where parameters $\theta$ are the DNN weights, $\mathbf{s}$ are the states, and $\mathbf{a}$ are the actions.
A Deep Deterministic Policy Gradient (DDPG) algorithm, introduced by Lillicrap et. al \cite{lillicrap2015continuous} to operate in continuous spaces, employs an actor-critic architecture and learns deterministic policies and expands them into continuous action spaces. Policy gradient methods directly optimize the policy \cite{wang2022deep}. Let the expected return be the performance measure of the policy $\pi_\theta$. Then 

\begin{equation}
    J(\theta) = V_{\pi_\theta}(s) = E_{\pi_\theta(s)}[\sum_a{Q(s,a)\pi_\theta(a|s)}].
\end{equation}

By differentiating $J(\theta)$ with respect to $\theta$, policy optimization is obtained as follows

\begin{equation}
    \nabla_\theta J(\theta) = E_{\pi_\theta(s)}[\sum_a{Q(s,a)\nabla_\theta \pi_\theta(a|s)}],
\end{equation}
$\theta \leftarrow \theta + \alpha \nabla_\theta J$, where $\alpha$ is the step size.

Based on the total predicted load of the distribution network, the EV charging pricing strategy uses DDPG algorithm to set the dynamic prices. The DDPG agent uses the actor-critic architecture to learn the Q-function of the EV charging pricing model. The policy is defined by the actor network, which determines the actions taken in a given state. Using the reward, the critic network evaluates the policy, which is updated continuously to maximize the reward. 

The next section shows the methodology used, which uses these techniques to determine the dynamic EV charging prices for the improvement of the grid utlilization.

\section{Distribution network load forecasting and dynamic EV charging pricing}
\label{sec:method}

\begin{figure*} [t]
    \centering
    \includegraphics[width=\textwidth]{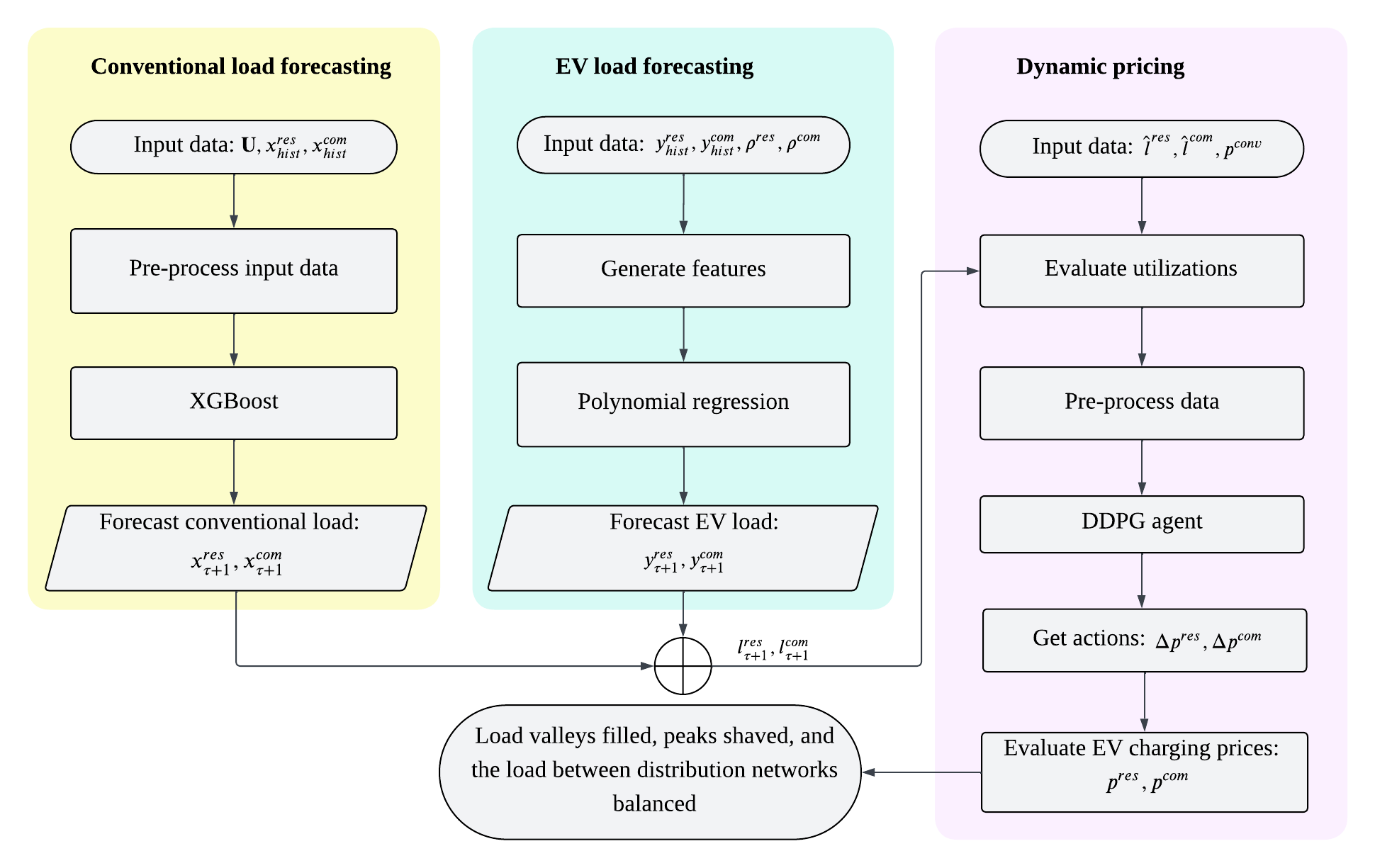}
    \caption{Distribution network load forecasting and dynamic EV charging pricing}
    \label{fig: forecast-pricing}
\end{figure*}

In smart power grids, the pattern of total power consumption by the grid load is used to determine the electricity Time-of-Use (ToU) pricing. With the integration of EVs in the grid, the load can be broadly classified into conventional load (home appliances, machinery, etc.) and the EV load.
In this work, we propose a dynamic EV charging pricing strategy that encourages an optimally distributed power consumption in a distribution network throughout the day, as well as balanced consumption between neighbouring distribution networks. We consider two distinct patterns of consumption in the distribution networks: the residential and commercial power consumption patterns.

We conduct the distribution network level day-ahead load forecasting and the dynamic EV charging pricing. As shown in Figure \ref{fig: forecast-pricing}, the proposed methodology is divided into three different stages. In the first stage, we conduct conventional load power consumption forecasting. The required input data consists of the historical environmental variables $\mathbf{U}$, and the associated residential load power consumption $\mathbf{x}_{hist}^{res}$ or commercial load power consumption $\mathbf{x}_{hist}^{com}$. Data is then pre-processed before being passed to the XGBoost algorithm for day-ahead predictions $\mathbf{x}_{\tau+1}^{res}$ and $\mathbf{x}_{\tau+1}^{com}$.

The second stage consists of EV load power consumption forecasting, which takes the historical EV power consumption load in the residential area $\mathbf{y}_{hist}^{res}$, historical EV power consumption load in the commercial area $\mathbf{y}_{hist}^{com}$ and the custom distribution of EVs in each area $\rho^{res}$, $\rho^{com}$ and generates the additional features before using polynomial regression to forecast day-ahead EV load consumption in each area $\mathbf{y}_{\tau+1}^{res}$ and $\mathbf{y}_{\tau+1}^{com}$. The day-ahead total power consumption in each distribution network is then used to set EV charging prices to accomplish peak-shaving, valley-filling, and load balancing between distribution networks.
The central pricing strategy sets the EV charging price for a distribution network, based on the power consumption pattern of an area where the distribution network is located.



\subsection{Distribution network conventional load forecasting} \label{subsec: conv_load}

The conventional load varies with time and has a predictable pattern characteristic to the location of the distribution network, whether in a residential or commercial area. It is highly impacted by environmental conditions.
To forecast the conventional load, we use the conventional load dataset that was collected from a distribution network in Morocco, freely accessible at \cite{morocco_data}. Its description is given as follows.

\begin{figure} [h!]
    \centering
    \includegraphics[width=0.49\textwidth]{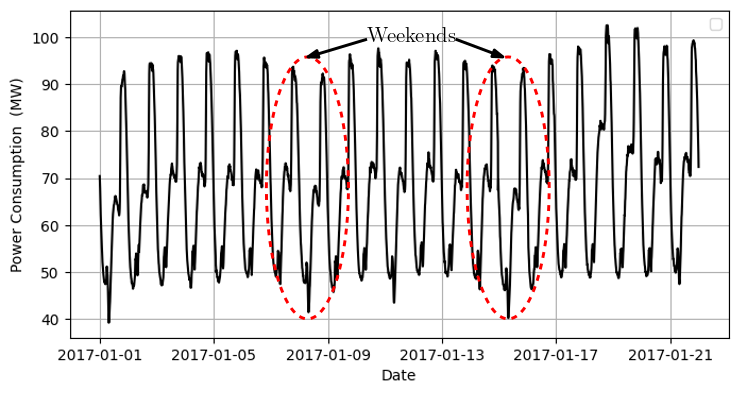}
    \caption{Total power consumption for 21 days}
    \label{fig: consump_21}
\end{figure}

\subsubsection{Conventional load data description and pre-processing}

The data was collected from Tetouan city in Morocco, from a distribution network powered by three zone stations. 
The load data was collected in 10-minute intervals from 01 January 2017 to 31 December 2017. Together with the power consumption, the environmental conditions variables shown in Table \ref{pccs_table} were also collected. The proposed dynamic EV charging pricing strategy in this work is in hourly intervals, we therefore further process the data by downsampling it to be in hourly intervals. In our downsampling process, we sum up power consumption data for each hour, and determine the mean for the environmental factors variables. Figure \ref{fig: consump_21} shows the plot of distribution network-level total power consumption for the first twenty one days.

From Figure \ref{fig: consump_21}, we observe that the total power consumption has a visible cycle that happens every seven days, such that week and weekend days' can be seen by the difference in consumption. For example, the highest power consumption peaks are realized during the weekdays, and the lowest valleys are realized on the weekends. Moreover, an observation on individual days' consumptions shows that peak daily consumption happens in the evening hours, while the least consumption happens at night hours, and fair consumption happens during the day. 
Based on these observations, we derive \textit{hour} and \textit{day} features based on the \textit{datetime} attribute in the dataset.

To determine how each feature contributes to the prediction of \textit{total power consumption} as a response variable, we determine the Pearson Correlation Coefficient (PCC) between the features and the response variable. Table \ref{pccs_table} shows the features and the corresponding PCC values.

\begin{table}
\centering
\caption{Table of features and the corresponding PCC values}
\label{pccs_table}
\begin{tblr}{
  width = \linewidth,
  colspec = {Q[317]Q[154]Q[454]},
  hline{1-2} = {-}{},
}
Feature           & Data type & Pearson correlation coefficient \\
Datetime            & Datetime  &                                 \\
Hour                & Integer   & 0.684575                        \\
Temperature         & Float     & 0.49178                         \\
WindSpeed           & Float     & 0.22419                         \\
GeneralDiffuseFlows & Float     & 0.153346                        \\
DiffuseFlows        & Float     & 0.036845                        \\
Day                 & Integer   & -0.064649                       \\
Humidity            & Float     & -0.302248                       
\end{tblr}
\end{table}

To prepare the data for machine learning-based prediction, we apply one-hot encoding to the categorical features: \textit{Day} and \textit{Hour}, and standardize the rest of the features using min-max scaling \cite{minmax}, given by \eqref{eq: minmaxnormalization}.

\begin{equation} \label{eq: minmaxnormalization}
 f(u_i) =\frac{u_i-min(\mathbf{u})}{max(\mathbf{u})-min(\mathbf{u})},
\end{equation}

where $\mathbf{u}$ is a feature vector. After the preparation of data, the three algorithms namely: polynomial regression,  Extreme Gradient Boosting (XGBoost), and neural network are used for the predictions. We use grid search to find the best combination of hyperparameters.

\subsection{Charging load of electric vehicles} \label{subsec: ev_load}

\begin{figure} [h]
    \centering
    \includegraphics[width=0.5\textwidth]{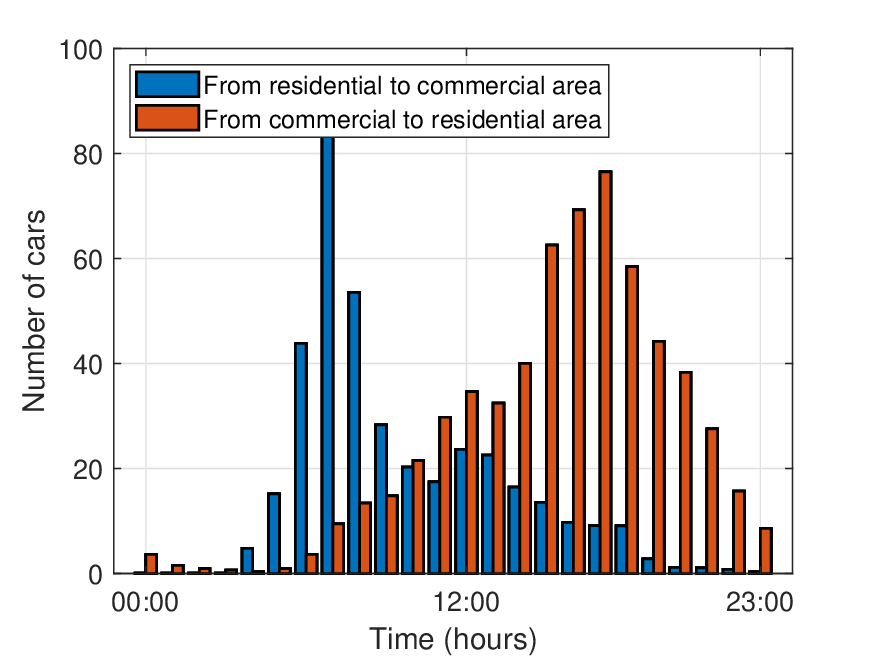}
    \caption{Vehicles' travel behaviour between residential and commercial regions based on U.S. NHTS data}
    \label{fig: commuting_behaviour}
\end{figure}

Research conducted by \cite{kim2019insights,sohail2023impact,gilleran2021impact} shows that the charging demand of EVs peaks when the electricity price is relatively low and it drops when the electricity price is relatively high. In \cite{kim2019insights}, it was established that in the presence of EVs, two electricity demand peaks in the residential area are formed. The first peak which is between 18:00 hours and 20:00 hours results from the usual household load, while the midnight peak comes from the charging of electric vehicles. In \cite{gilleran2021impact}, it was established that in the presence of EVs, the retail buildings' load peaks between 12:00 hours and 18:00 hours when the load of the building is also highest since the EV users prefer to charge their EVs while doing the shopping. The charging demand drops between 20:00 hours and 06:00 hours.
Analysing the typical travel behaviour of vehicles commuting between residential and commercial areas in a day on the NHTS dataset \cite{nhts_website}, the bar chart of vehicles commuting between residential and commercial areas takes the shape shown in Figure \ref{fig: commuting_behaviour}.

Based on the commuting behaviour in Figure \ref{fig: commuting_behaviour}, we do the custom distribution that approximates the distribution of EVs in the residential and commercial areas. The NHTS data show the general travel behaviour that can be applied elsewhere. For example, people leave home in the residential areas to go to work, school, etc, in the morning hours, and are mostly back at home in the late afternoons and evenings. Our model assumes that EV users can commute between residential and commercial areas at a given time.
Authors in \cite{kuang2024unravelling} further show that EV load can be further traced in the nearby areas if the EV charging pricing strategy for one area is varied, therefore we further formulate our model based on an assumption that for an EV user to choose to charge in a certain area, the willingness to charge is proportional to the price variance between the distribution networks and the EV's location at the given time.
In the residential area, we define the EVs' charging demand inspired by the users' willingness to charge as follows.

Let $t \in \{1,\dots,24\}$ index the hour of the day. We denote by $\rho^{res}_t\geq 0$ the hourly probability distribution for residential EV charging with $\sum_{t=1}^{24}\rho^{res}_t=1$, and by $g^{res}$ the total residential EV energy demand per day (kWh), approximated by a Gaussian variable with mean $\mu^{res}$ and variance $\sigma^{res}$.
Residential and commercial prices at hour $t$ are $p^{res}_t$ and $p^{com}_t$ (currency/kWh).
The residential charging demand is mainly influenced by the following two components: (A) normalized price sensitivity, and (B) relative price difference. Their hourly contributions are
\begin{align}
y^{res}_{A,t} &= c_1^{res}\, \rho^{res}_t\, g^{res}\left(1-\frac{p^{res}_t}{p_{tot}}\right), \label{eq: D_res_prop_1}\\
y^{res}_{B,t} &= c_2^{res}\, \rho^{res}_t\, g^{res}\left(p^{com}_t-p^{res}_t\right), \label{eq: D_res_prop_2}
\end{align}
where $c_1^{res},c_2^{res}\geq 0$ are parameters with units chosen to yield demand in kWh, and $p_{tot}$ is the sum of unique charging prices. 

The total residential charging demand is then expressed as a combination of these components shown in Equation \eqref{eq: D_res_props_comb}.
\begin{multline} \label{eq: D_res_props_comb}
y^{res}_t = \rho^{res}_t\, g^{res}\Big[\kappa^{res}\, c_1^{res}\Big(1-\frac{p^{res}_t}{p_{tot}}\Big) \\
+ (1-\kappa^{res})\, c_2^{res}\big(p^{com}_t-p^{res}_t\big)\Big], \quad \kappa^{res} \in [0,1].
\end{multline}
The charging demand of EVs in the commercial area distribution network,
derived in a similar manner, is given by Equation \eqref{eq: D_com_props_comb}.
\begin{multline} \label{eq: D_com_props_comb}
y^{com}_t = \rho^{com}_t\, g^{com}\Big[\kappa^{com}\, c_1^{com}\Big(1-\frac{p^{com}_t}{p_{tot}}\Big) \\
+ (1-\kappa^{com})\, c_2^{com}\big(p^{res}_t-p^{com}_t\big)\Big], \quad \kappa^{com} \in [0,1].
\end{multline}

Given the dynamics of the charging requirements of the EV users, the values of $c^{res}_1$, $c^{res}_2$, $c^{com}_1$, $c^{com}_2$, $g^{res}$, and $g^{com}$ may vary from time to time. 
To make future predictions for $\mathbf{y}^{res}$, Equation \eqref{eq: D_res_props_comb} and \eqref{eq: D_com_props_comb} can be approximated with a learning algorithm. The structure of these equations is suited for a polynomial regression with $\mathbf{y}^{res}$ and $\mathbf{y}^{com}$ as response variables (outputs to be predicted) and the prediction features derived from distribution probability of the EVs and the charging prices at the given time in a distribution network.

\subsection{Dynamic pricing}

Let $x_t$ and $y_t$ be the conventional load and the EV charging load in a distribution network at time $t$, respectively. The total residential ($l^{res}_t$) and commercial ($l^{com}_t$) distribution networks loads are given by Equations \eqref{eq: res_load} and \eqref{eq: com_load}.

\begin{equation} \label{eq: res_load}
    l^{res}_t = x^{res}_t + y^{res}_t \leq \hat{l}^{res}, 
\end{equation}

and 

\begin{equation} \label{eq: com_load}
    l^{com}_t = x^{com}_t + y^{com}_t \leq \hat{l}^{com}, 
\end{equation}

where $x^{res}_t$, $y^{res}_t$, $x^{com}_t$, and $y^{com}_t$ can be forecast as shown in \ref{subsec: conv_load} and \ref{subsec: ev_load}. For the dynamic pricing that follows, we introduce the constraints $\hat{l}^{res}$ and $\hat{l}^{com}$, which are the maximum loads that can be handled by the residential and commercial distribution networks respectively.

The EV charging price should be set in such a way that the following two conditions are satisfied: first, for peak-shaving and valley-filling, EV charging price must be lowest at the time when the distribution network load is at its minimum, and highest when the distribution network load is at its peak. This will ensure that more EVs are encouraged to charge when the total distribution network load is at minimum, and discouraged to charge when it is at peak, thereby causing the distribution network load to be uniform over time.
Secondly, the EV charging prices should attract EVs to a relatively less utilized distribution network and offload a highly utilized distribution network at a given time, to balance the load between the distribution networks in close vicinity, thereby ensuring optimal usage of resources among the neighbouring distribution networks.

That is, given a conventional electricity price $p_t^{conv}$, EV charging price in a residential distribution network $p_t^{res}$ is given by \eqref{eq: p_res}:

\begin{equation} \label{eq: p_res}
    p_t^{res} = p_t^{conv} + \Delta p_t^{res}, 
\end{equation}

where

\begin{equation}
    \Delta p_t^{res} = \Delta p^{res}_{a,t} + \Delta p^{res}_{b,t}.
\end{equation}

$\Delta p^{res}_{a,t}$ and $\Delta p^{res}_{b,t}$ are components for satisfying the first and second EV charging pricing conditions, respectively. For the first condition to be satisfied, $\Delta p^{res}_{a}$ must follow the load profile $l^{res}$, and $\Delta p^{res}_b$ must follow the difference in neighboring distribution networks' utilizations. This leads to the requirement to solve \eqref{eq: p_a} and \eqref{eq: p_b}.

\begin{equation} \label{eq: p_a}
    \min_{\Delta p^{res}_{a,t}} |\Delta p^{res}_{a,t} - c^{res}_al_t^{res}|, 
\end{equation}

\begin{equation} \label{eq: p_b}
    \min |\Delta p^{res}_{b,t} - c^{res}_b\left(\frac{l_t^{res}}{\hat{l}^{res}} - \frac{l_t^{com}}{\hat{l}^{com}}\right)|,
\end{equation}

such that $\Delta p^{res}_{a,t} \geq 0$, $\Delta p^{res}_{b,t} \geq 0$, and if $\frac{l_t^{res}}{\hat{l}^{res}} \leq \frac{l_t^{com}}{\hat{l}^{com}}$, then $\Delta p^{res}_{b,t} = 0$.
$c^{res}_a, c^{res}_b \geq 0$ ensure proportionality between $\Delta p^{res}_a$ and $l^{res}$, and $\Delta p^{res}_b$ and $\left(\frac{l^{res}}{\hat{l}^{res}} - \frac{l^{com}}{\hat{l}^{com}}\right)$ respectively.






The dynamic prices set for the current load can affect future load patterns. To ensure high adaptability to changing distribution network load and utilization of the distribution networks in the EV charging pricing strategy, a method that can adapt continuously to changing environments is required. We therefore formulate the pricing problem as a Markov Decision Process (MDP) and solve equations \eqref{eq: p_res}-\eqref{eq: p_b} using the deep reinforcement learning (DRL) approach. The MDP is detailed as follows. The environment state for the residential area distribution network is the continuous load $l^{res}$, sampled at an hourly interval.

The actions taken by the residential area agent are $\Delta p^{res}$, whereby $\Delta p^{res}_a$ and $\Delta p^{res}_b$ are evaluated as in \eqref{eq: a1_res} and \eqref{eq: a2_res}.

\begin{equation} \label{eq: a1_res}
    \Delta p^{res}_a = c^{res}_al^{res}, 
\end{equation}

\begin{equation} 
\label{eq: a2_res}
\Delta p^{res}_b = 
    \begin{cases}
        c^{res}_b\left(\frac{l^{res}}{\hat{l}^{res}} - \frac{l^{com}}{\hat{l}^{com}}\right), & \text{if }  \frac{l^{res}}{\hat{l}^{res}} > \frac{l^{com}}{\hat{l}^{com}} \\
        0, & \text{otherwise}.
    \end{cases}
\end{equation}




The state transition probability is influenced by different factors that lead to the deviations in load. To simulate the real world environment, we consider that it is unknown. The reward function for the residential area agent $R^{res}$ is designed as in \eqref{eq: res_reward}.


\begin{strip}
    \begin{equation} \label{eq: res_reward}
        R^{res} = 
        \begin{cases}
            -\omega^{res}_1|\Delta p^{res}-c^{res}_al^{res}| - \omega^{res}_2 |
            \left[\Delta p^{res} - c^{res}_b\left(\frac{l^{res}}{\hat{l}^{res}} - \frac{l^{com}}{\hat{l}^{com}}\right)\right]|
         , & \text{if }  \frac{l^{res}}{\hat{l}^{res}} > \frac{l^{com}}{\hat{l}^{com}} \\
            -|\Delta p^{res}-c^{res}_al^{res}|, & \text{otherwise}.
        \end{cases}
    \end{equation}
\end{strip}


The commercial area pricing strategy is also done in the same manner as the residential. We test the performance of the following three widely used DRL algorithms: DDPG, SAC and PPO, and show the effect of the dynamic prices set by the best-performing algorithm. For the effective training of the actor-critic networks, the load $l^{res}$ is standardized between $0$ and $1$ using min-max scaling.
The utilization difference $\frac{l^{com}}{\hat{l}^{com}} - \frac{l^{res}}{\hat{l}^{res}}$ is also standardized between $0$ and $1$, resulting in both $c^{res}_a$ and $c^{res}_b$ set to unit values in the calculation of the reward. The resulting actions $\Delta p^{res}$ can then be scaled to the desired values by the distribution network operator. Similar steps are taken for the commercial area agent.

Table \ref{drl_parameters} shows the training parameters for the tested DRL algorithms. For fair performance comparison, the same parameter values were set for common parameters, and the algorithms were trained on the same dataset, for $15,000$ epochs. The values of the unique parameters for each algorithm have also been shown. For each algorithm, the rewards and the actor and critic losses were recorded. We compare the rewards for all the three algorithms, and the actor and critic losses for DDPG and SAC. The PPO actor and critic losses are calculated in a manner that is different from DDPG and SAC because of the PPO's on-policy fundamental property, while DDPG and SAC are off-policy algorithms. The results are shown in the next section.

\begin{table}
\centering
\caption{DRL parameters}
\label{drl_parameters}
\begin{tblr}{
  width = \linewidth,
  colspec = {Q[315]Q[419]Q[185]},
  cell{2}{1} = {r=5}{},
  hline{1-2} = {-}{},
}
\textbf{Algorithm}   & \textbf{Parameter}        & \textbf{Values} \\
All three algorithms & Actor network layer size  & (64, 64,24)     \\
                     & Critic network layer size & (64, 64, 1)     \\
                   & Learning rate             & 0.0003          \\
                     & Neural networks optimizer & Adam            \\
                     & Discount factor           & 0.99            \\
                     & & \\
DDPG                 & Soft update rate          & 0.005           \\
                     & & \\
SAC                  & Soft update rate          & 0.005           \\
                     & Automatic entropy tuning  & TRUE            \\
                     & Temperature coefficient   & 0.2             \\
                     & & \\
PPO                  & Genereralized advantage   & 0.95            \\
                     & estimate bias-variance    &                 \\
                     & PPO clipping parameter    & 0.2             \\
                     & Policy epochs             & 4               \\
                     & Entropy coefficient       & 0.001           \\
                     & Gradient clipping         & 0.5             
\end{tblr}
\end{table}

\section{Results and Discussion}
\label{sec:discussion}

In this section, we present and discuss the results obtained when testing the proposed dynamic EV charging pricing method. We first present conventional load prediction results, followed by day-ahead EV load forecasting results, and then finally EV charging pricing results, where a comparison between conventional ToU pricing, dynamic pricing for peak-shaving and valley-filling (PV EV charging pricing), and dynamic pricing for peak-shaving, valley-filling, and inter-distribution networks load balancing (PVB EV charging pricing) is made.
All simulations are done using the Python programming language, and run on a Visual Studio Code environment. The codebase can be found at \url{https://github.com/Leloko/DRLDynamicPricing}.

\subsection{Conventional load prediction}

For forecasting the conventional load using the environmental variables, the performances of the following regression algorithms are compared: polynomial regression, XGBoost, and the feed-forward neural network. Grid search is used to find the best combination of parameters for each algorithm. Table \ref{reg_algos_table} shows the regression algorithms tested and the best combinations of parameters that yield the best results.

\begin{table}[h!]
\centering
\caption{Tested regression algorithms and the best combinations of parameters}
\label{reg_algos_table}
\begin{tblr}{
  width = \linewidth,
  colspec = {Q[160]Q[250]Q[160]Q[160]},
  hline{1-2} = {-}{},
}
Algorithm             & Parameter         & Tested Values                          & Best values                         \\
Neural network        & Hidden layer size & 50, 100, (50, 50), (16, 8), (16, 8, 4) & (16, 8, 4)   \\
                      & L2 regularization term             & 0.0001, 0.001                          & 0.001                                        \\
                      & Learning rate     & constant, adaptive                     & constant                           \\
                      &                   &                                        &                                              \\
XGBoost               & Maximum number of boosting trees   & 50, 70, 100, 150, 200                  & 200          \\
                      & Maximun depth     & 3, 5, 7                                & 5                                            \\
                      & Learning rate     & 0.01, 0.1, 0.2                         & 0.2                                          \\
                      &                   &                                        &                                              \\
Polynomial regression & Degree            & 2                                      & 2            

\end{tblr}
\end{table}

\begin{figure} [h]
    \centering
    \includegraphics[width=0.5\textwidth]{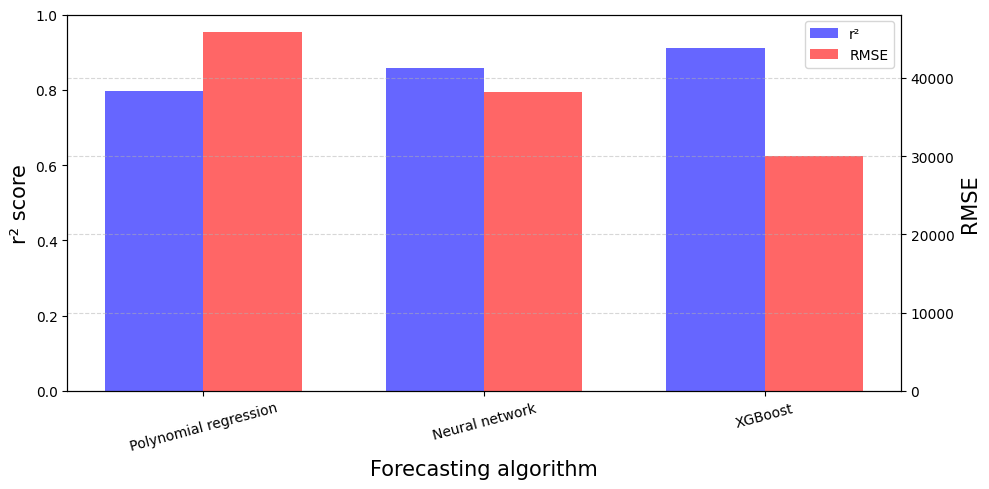}
    \caption{RMSE and r\^{}2 results of conventional load forecasting regression algorithms}
    \label{fig: conv_forecasting}
\end{figure}

Figure \ref{fig: conv_forecasting} shows the regression algorithms tested and the results obtained with the best combinations of parameters. We use Root Mean Squared Error (RMSE) and R-squared ($r^2$) for the performance comparison. The lower the RMSE, and the closer the $r^2$ to $1$, the better the performance.
As shown in Figure \ref{fig: conv_forecasting}, XGBoost outperforms polynomial regression and neural networks, with the least RMSE of 300082.66 and the best $r^2$ of 0.912.

\subsection{Electric vehicles load prediction}

\begin{figure} [h!]
    \centering
     \begin{subfigure}{0.24\textwidth}
         \includegraphics[width=\textwidth]{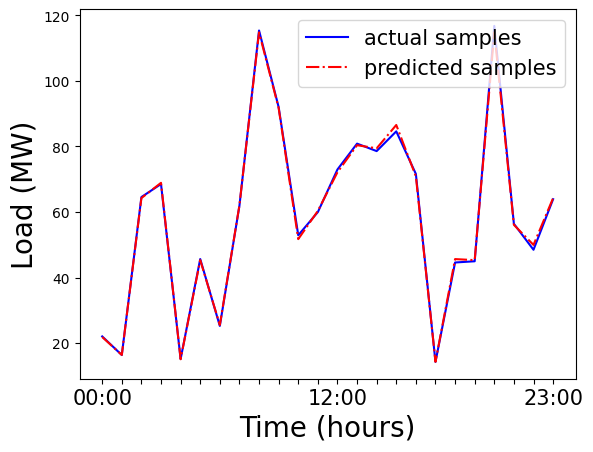}
         \caption{Residential area}
         \label{fig: pred_res_demand}
    \end{subfigure}
    \begin{subfigure}{0.24\textwidth}
         \includegraphics[width=\textwidth]{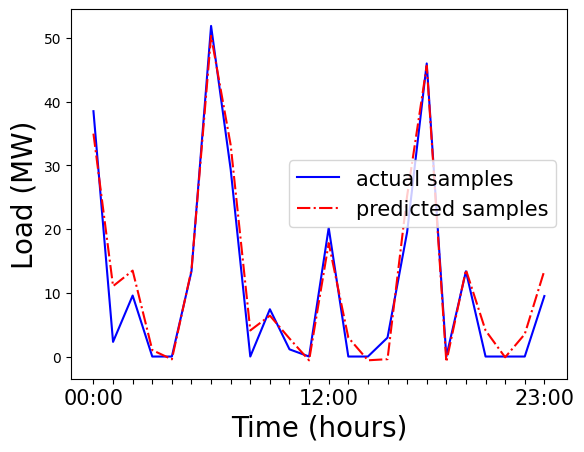}
    \caption{Commercial area}
    \label{fig: pred_com_demand}
    \end{subfigure}
    \caption{Predicted EV load demand}
    \label{fig: pred_res_com_demand}
\end{figure}

The hourly EV charging demand data of 365 days generated using equations \eqref{eq: D_res_props_comb} and \eqref{eq: D_com_props_comb} is used to train and validate the second-order polynomial regression algorithm. 80\% of the data was used for training and 20\% for validation. Figure \ref{fig: pred_res_com_demand} shows the predicted and the actual samples for a single day in residential and commercial areas.

For residential area prediction $r^2 = 0.999$, $\text{RMSE} = 1101.75$, and $r^2 = 0.957$, $\text{RMSE} = 3567.38$ for the commercial area prediction. The values of both the $r^2$ and RMSE indicate good performance for both the residential and commercial areas.

\subsection{Electric vehicles' charging dynamic pricing}

\begin{figure} [h!]
     \begin{subfigure}{0.24\textwidth}
         \includegraphics[width=\textwidth]{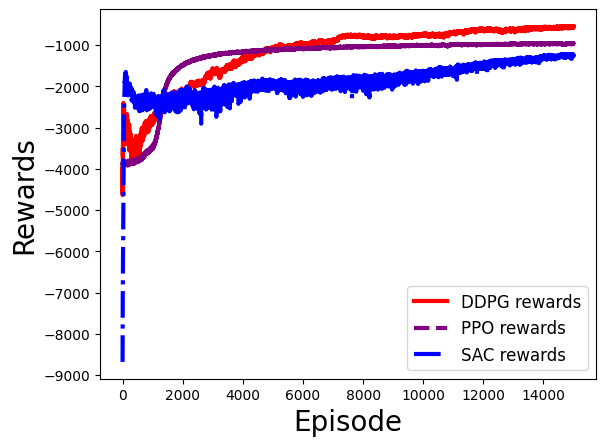}
         \caption{Rewards per episode}
         \label{fig: drl_rewards}
    \end{subfigure}
    \begin{subfigure}{0.24\textwidth}
         \includegraphics[width=\textwidth]{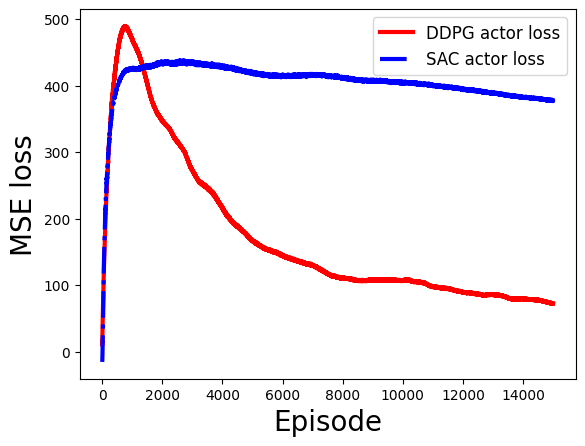}
    \caption{Actor losses}
    \label{fig: drl_actor_losses}
    \end{subfigure}
    \begin{subfigure}{0.24\textwidth}
         \includegraphics[width=\textwidth]{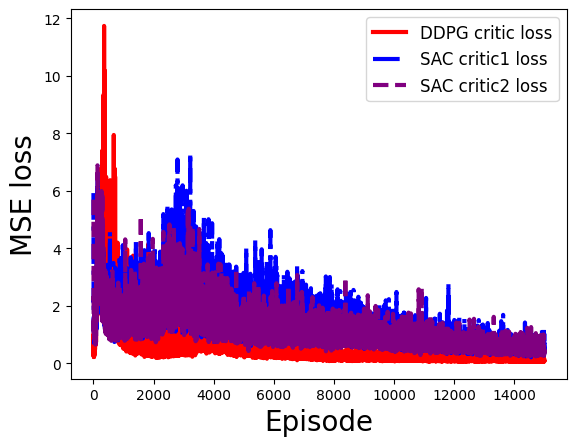}
    \caption{Critic losses}
    \label{fig: drl_critic_losses}
    \end{subfigure}
    
    \caption{Performance comparison between different DRL algorithms}
    \label{fig: drl_comparisons}
\end{figure}

The performance comparison of DDPG, SAC, and PPO algorithms is shown in Figure \ref{fig: drl_comparisons}. In Figure \ref{fig: drl_rewards}, the plot of the rewards over the $15,000$ episodes is shown. For all algorithms, the rewards improve over the iterations, until they converge to their best value. As shown in Figure \ref{fig: drl_rewards}, DDPG outperformed both SAC and PPO interms of the best rewards values. In Figures \ref{fig: drl_actor_losses} and \ref{fig: drl_critic_losses}, actor and critic networks' MSE losses for DDPG and SAC are compared.
Since the losses for PPO algorithm are not directly comparable with DDPG and SAC losses, we do not show them in these plots.
In both cases, DDPG losses converge to better MSE values than SAC.

These results show the superiority of DDPG over SAC and PPO for this task. We therefore use DDPG for the EV dynamic pricing results that follow.
The dynamic EV charging prices incorporate the conventional pricing. For our simulation, we adopt the conventional prices that are shown in Table \ref{tou_prices}, inspired by \cite{khan2022effects}.

\begin{table}
\centering
\caption{Conventional ToU pricing structure}
    \label{tou_prices}
    \begin{tabular}{ll}
    \hline
    Time & Price\\
    \hline
    00:00 - 07:00 & 0.4 \\
    07:00 - 10:00 , 15:00 - 18:00, 21:00 - 0:00 & 0.7    \\
    10:00 - 15:00 , 18:00 - 21:00 & 1        \\
    \hline  
    \end{tabular}
\end{table}

\subsubsection{Conventional ToU pricing}

Before the incorporation of the proposed method, the major peak of the conventional load in the residential area is in the evening, while the EV load peaks when the electricity price is lowest at night. The result is high distribution network load in the evening and at night, and relatively low load during the day. In the commercial area, both the conventional load and the EV load peak during the day, leading to an increased gap between the peak and the valley of the total load when the EV charging is incorporated.


\subsubsection{Load peak-shaving and valley-filling}

\begin{figure} [h!]
    \centering
     \begin{subfigure}{0.24\textwidth}
         \includegraphics[width=\textwidth]{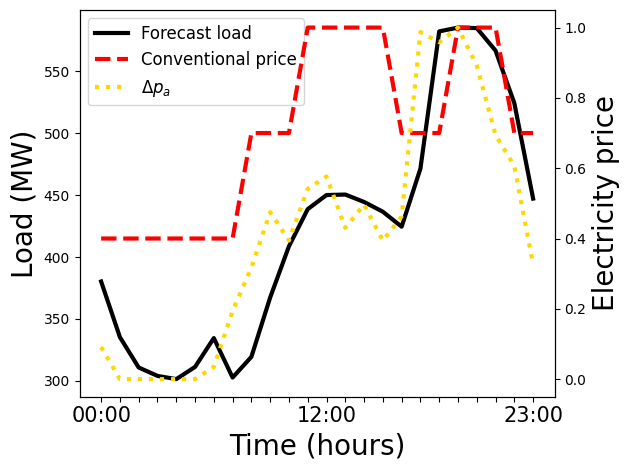}
         \caption{Residential area}
         \label{fig: pv_conv_res_load}
    \end{subfigure}
    \begin{subfigure}{0.24\textwidth}
         \includegraphics[width=\textwidth]{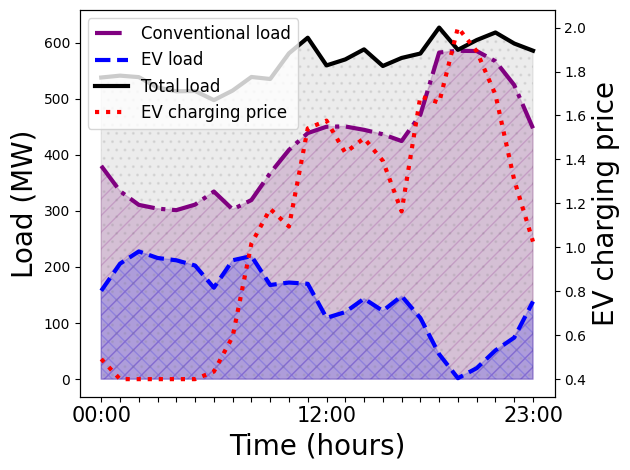}
         \caption{Residential area}
         \label{fig: pv_opt_res_load}
    \end{subfigure}
    \begin{subfigure}{0.24\textwidth}
         \includegraphics[width=\textwidth]{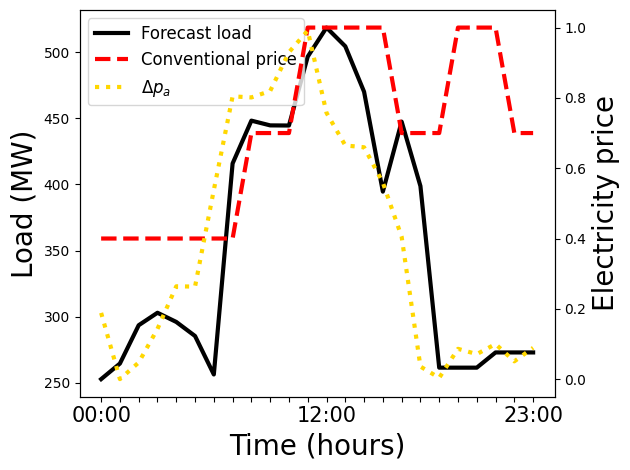}
        \caption{Commercial area}
        \label{fig: pv_conv_com_load}
    \end{subfigure}
    \begin{subfigure}{0.24\textwidth}
         \includegraphics[width=\textwidth]{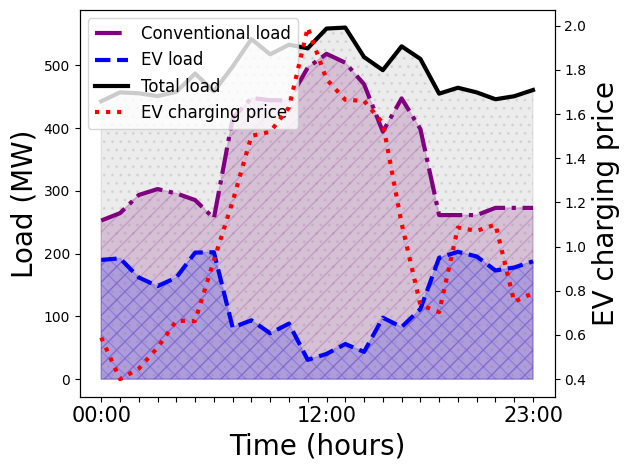}
        \caption{Commercial area}
        \label{fig: pv_opt_com_load}
    \end{subfigure}
    \caption{Actions for peak-shaving and valley-filling in different distribution networks}
    \label{fig: pv_opt_res_com_load}
\end{figure}

Figure \ref{fig: pv_opt_res_com_load} shows the forecast load, conventional electricity price ($p^{conv}$) and the actions taken for peak-shaving and valley-filling ($\Delta p_a$) in the residential and commercial areas' distribution networks, as well as the resulting impact of the modified EV charging pricing structure. 

In Figures \ref{fig: pv_conv_res_load} and \ref{fig: pv_conv_com_load}, the actions that are taken by the agents in the residential and commercial distribution networks are shown, respectively. $\Delta p_a$ follows the load in order to encourage charging of the EVs when the load is at the minimum, and discourage charging when the load is at peak.
Figures \ref{fig: pv_opt_res_load} and \ref{fig: pv_opt_com_load} show the impact of the modified EV charging price, which is the sum of conventional electricity price and $\Delta p_a$. While the conventional load stays the same, the EV load shifts according to the EV charging price, causing the total load to be uniform in both distribution networks.

\subsubsection{Inter-distribution network load-balancing}

\begin{figure} [h!]
    \centering
     \begin{subfigure}{0.24\textwidth}
         \includegraphics[width=\textwidth]{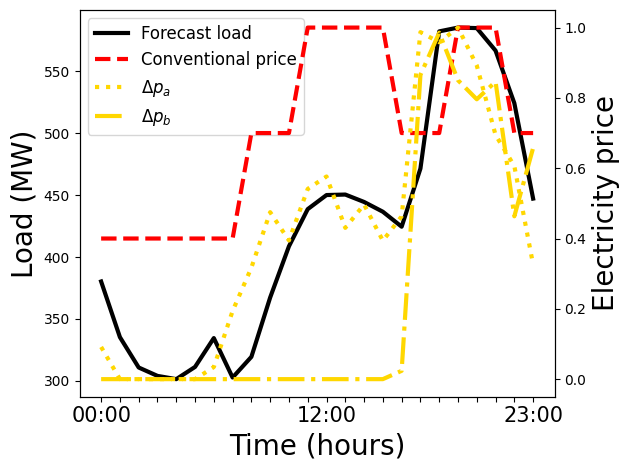}
         \caption{Residential area}
         \label{fig: pvb_conv_res_load}
    \end{subfigure}
    \begin{subfigure}{0.24\textwidth}
         \includegraphics[width=\textwidth]{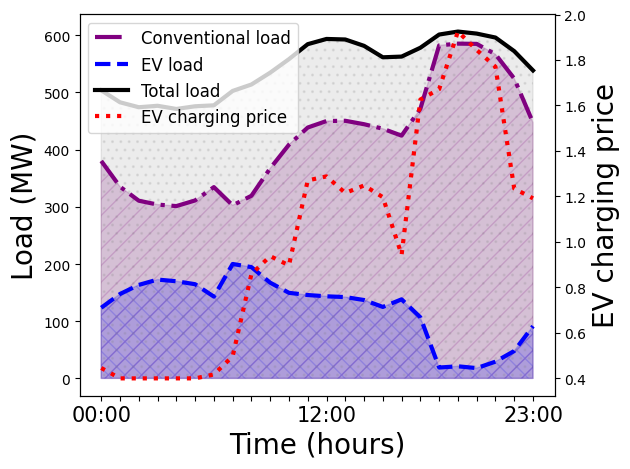}
         \caption{Residential area}
         \label{fig: pvb_opt_res_load}
    \end{subfigure}
    \begin{subfigure}{0.24\textwidth}
         \includegraphics[width=\textwidth]{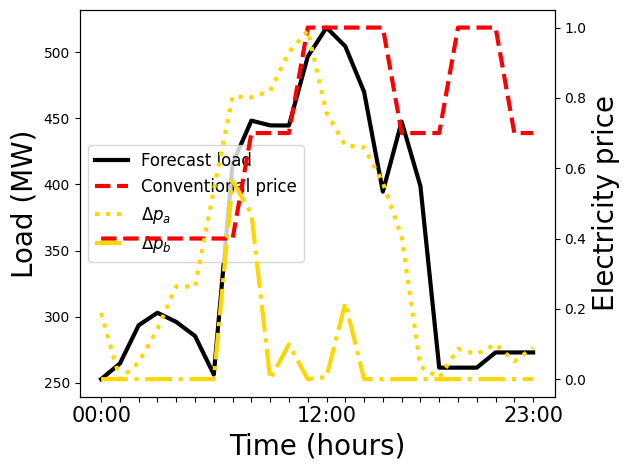}
        \caption{Commercial area}
        \label{fig: pvb_conv_com_load}
    \end{subfigure}
    \begin{subfigure}{0.24\textwidth}
         \includegraphics[width=\textwidth]{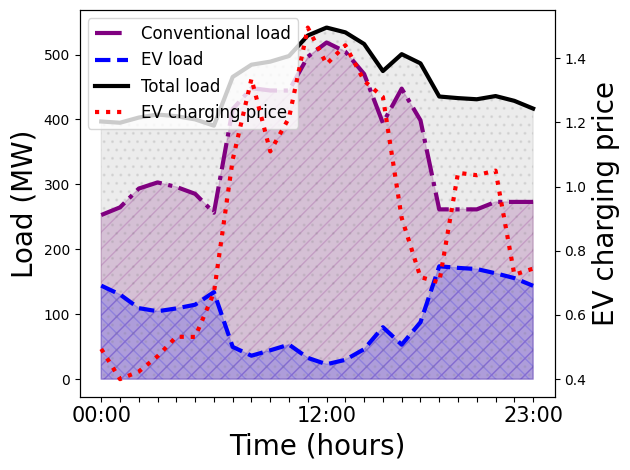}
        \caption{Commercial area}
        \label{fig: pvb_opt_com_load}
    \end{subfigure}
    \caption{Actions for peak-shaving, valley-filling and load-balancing in different distribution networks}
    \label{fig: pvb_opt_res_com_load}
\end{figure}

In Figure \ref{fig: pvb_opt_res_com_load}, the forecast load, conventional electricity price ($p^{conv}$) and the actions taken for peak-shaving, valley-filling, and inter-distribution network load balancing ($\Delta p_a$ and $\Delta p_b$) in the residential and commercial areas' distribution networks, as well as the resulting impact of the optimal EV charging pricing structure are shown.

In Figures \ref{fig: pvb_conv_res_load} and \ref{fig: pvb_conv_com_load}, the actions that are taken by the agents in the residential and commercial distribution networks are shown, respectively. Apart from $\Delta p_a$, $\Delta p_b$ is also shown, which are the actions taken to encourage offloading from a highly utilized distribution network to a relatively less utilized network.
Figures \ref{fig: pvb_opt_res_load} and \ref{fig: pvb_opt_com_load} show the impact of the optimal EV charging price, which is the sum of conventional electricity price and weighted $\Delta p_a$ and $\Delta p_b$. In this case, we chose equal priority for peak-shaving and valley-filling in each distribution network and inter-distribution network load-balancing and set $\omega_a = \omega_b = 0.5$. The result is shifted EV load causing the total load to be uniform across the different times of the day and to be balanced between the two distribution networks.

\subsubsection{The effect of different penetration levels of EVs in distribution networks under PVB EV charging pricing}

\begin{figure} [h!]
    \centering
     \begin{subfigure}{0.24\textwidth}
         \includegraphics[width=\textwidth]{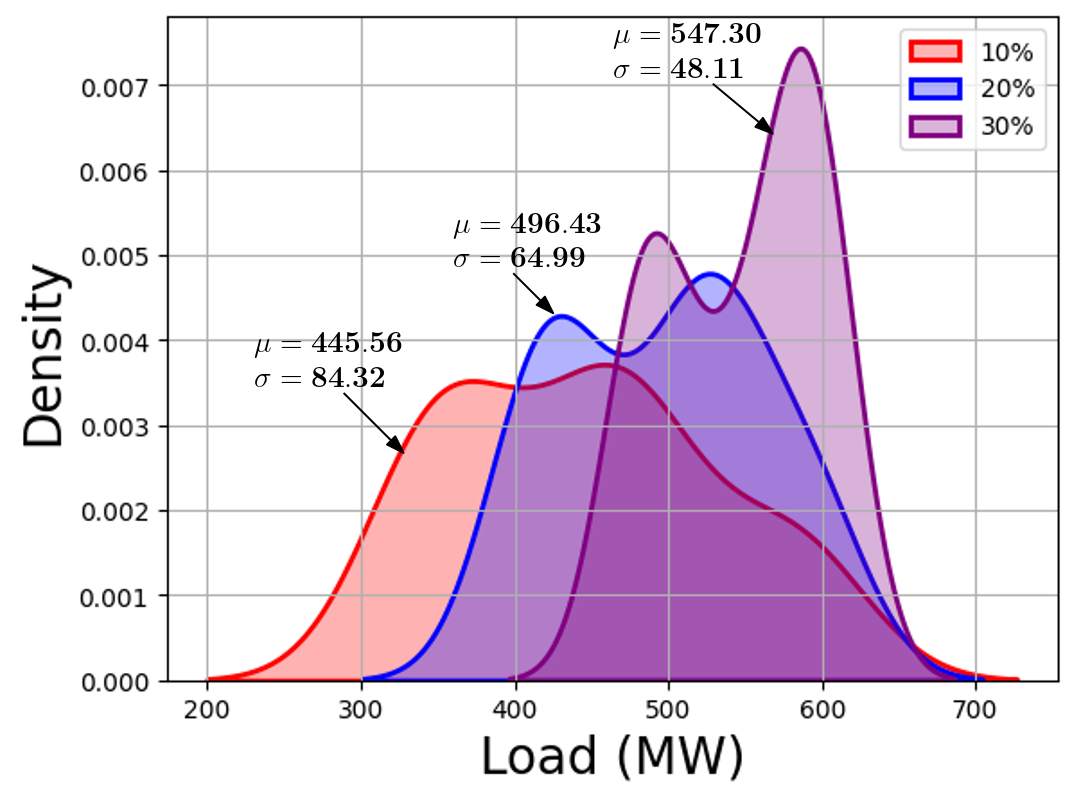}
         \caption{Residential area}
         \label{fig: evs_pen_res}
    \end{subfigure}
    \begin{subfigure}{0.24\textwidth}
         \includegraphics[width=\textwidth]{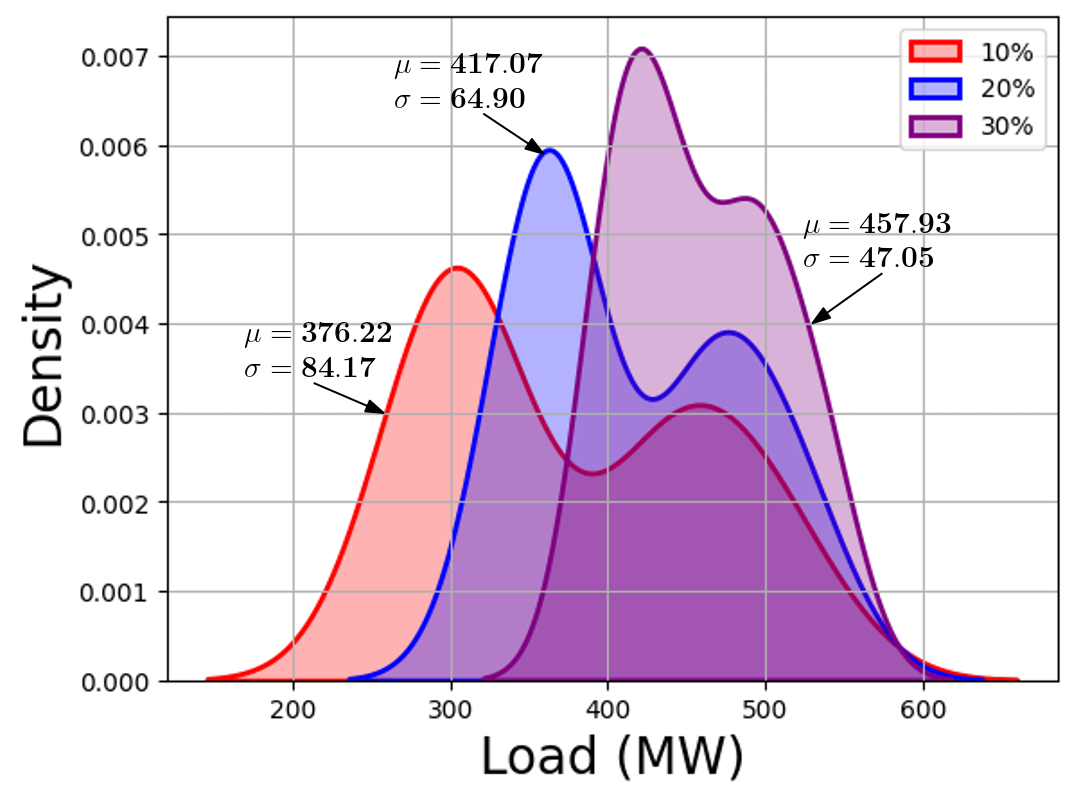}
        \caption{Commercial area}
        \label{fig: evs_pen_com}
    \end{subfigure}
    \caption{KDE plots of the distribution network-level load for different penetration levels of EVs under PVB EV charging pricing}
    \label{fig: evs_pen_res_com}
\end{figure}

Figure \ref{fig: evs_pen_res_com} shows the Kernel Density Estimation (KDE) plot of the distribution networks load under PVB EV charging pricing for different EV load penetrations. EV load contributions shown are in the quantities of 10\%, 20\%, and 30\% of the total distribution network load.

In both the commercial and residential areas distribution networks, the standard deviation ($\sigma$) is seen to decrease with the increase in EV load penetration, the lowest values of the load increase significantly, consequently causing the mean ($\mu$) values to increase, while the peak values do not change much. This is because the load peak values mostly come from the conventional load, and PVB EV charging pricing strategy encourages most charging of the EVs to happen at the load valleys. It is observed from Figure \ref{fig: evs_pen_res_com} that the higher the penetration of EVs under PVB EV charging pricing strategy, the more the uniform power consumption will be realized in the distribution networks.

\begin{figure} [h!]
    \centering
     \begin{subfigure}{0.24\textwidth}
         \includegraphics[width=\textwidth]{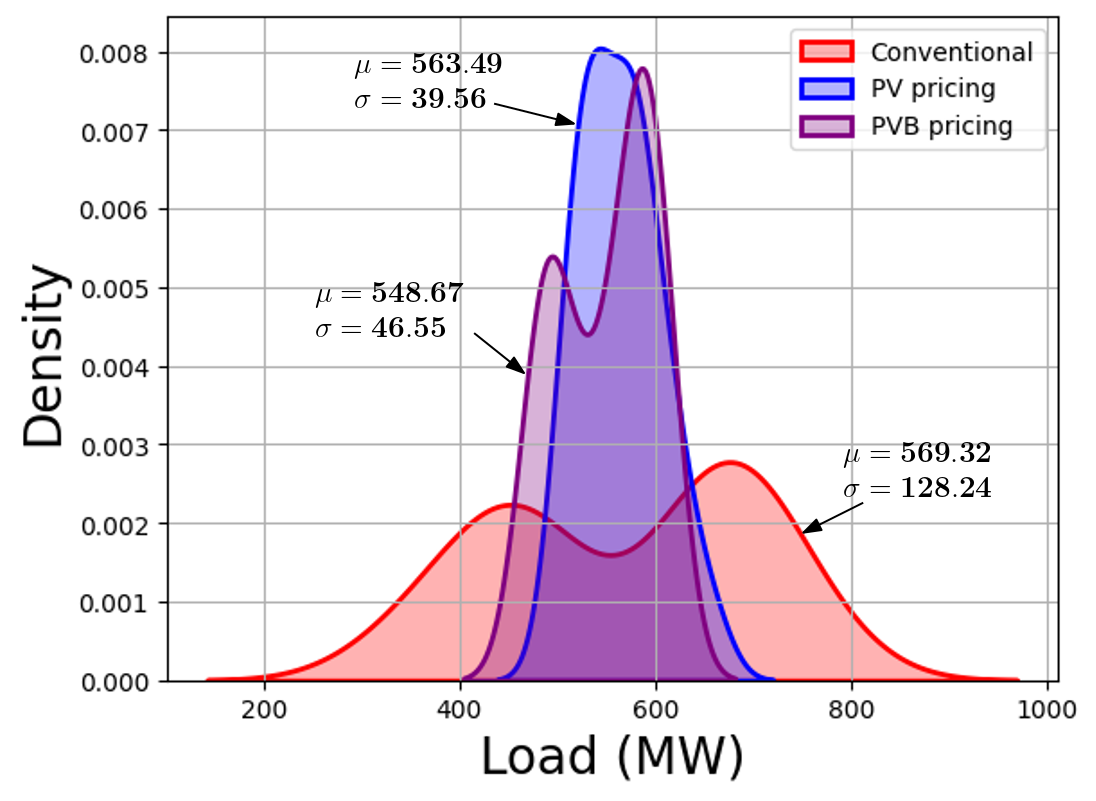}
         \caption{Residential area}
         \label{fig: kde_res}
    \end{subfigure}
    \begin{subfigure}{0.24\textwidth}
         \includegraphics[width=\textwidth]{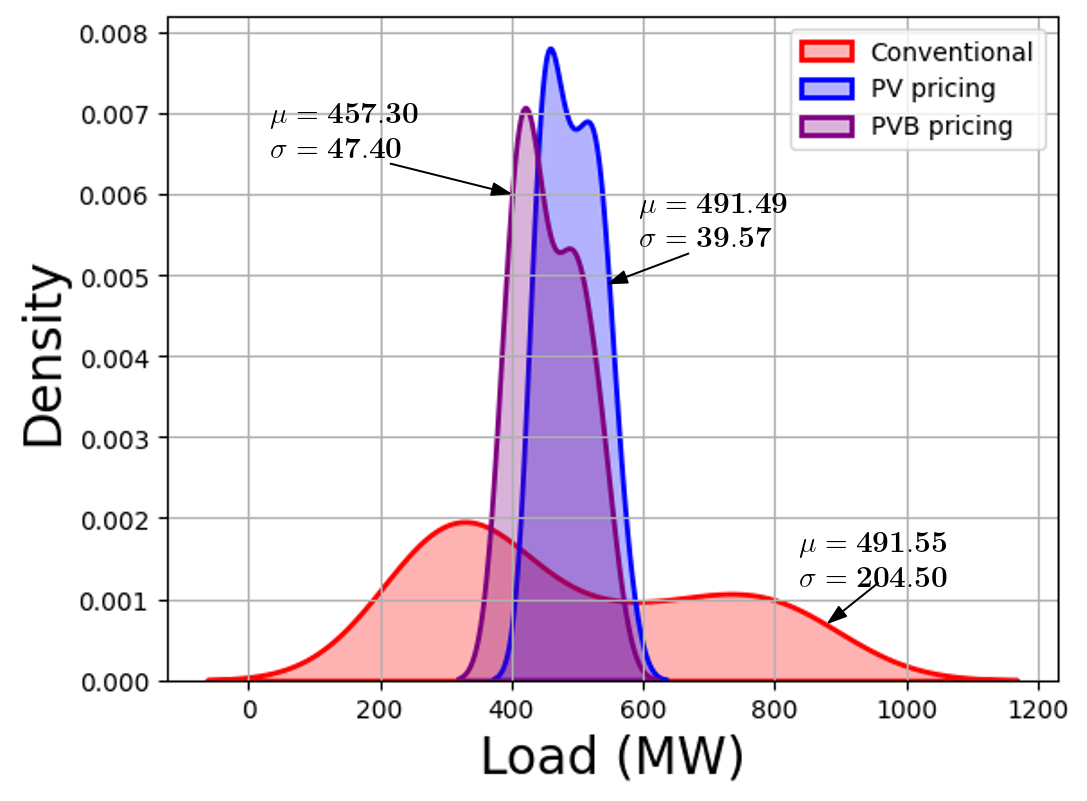}
        \caption{Commercial area}
        \label{fig: kde_com}
    \end{subfigure}
    \caption{KDE plots of distribution network-level power consumption under different EV charging pricing strategies}
    \label{fig: kde_res_com}
\end{figure} 

\subsubsection{Comparison of the effect of different EV charging pricing strategies on load distribution}
Using the 30\% EV penetration rate, the effect of the three EV charging pricing strategies: traditional ToU pricing strategy, PV EV charging pricing strategy and our proposed PVB EV charging pricing strategy, on the distribution of power consumption are shown in Figure \ref{fig: kde_res_com}.
As shown in Figure \ref{fig: kde_res_com}, in both residential and commercial areas distribution networks, traditional ToU pricing performs worst, while PV and PVB EV charging pricing strategies distribute the power consumption in the distribution network at the relatively close margins, resulting in a uniform load distribution throughout.

\begin{figure} [h!]
    \centering
     \begin{subfigure}{0.24\textwidth}
         \includegraphics[width=\textwidth]{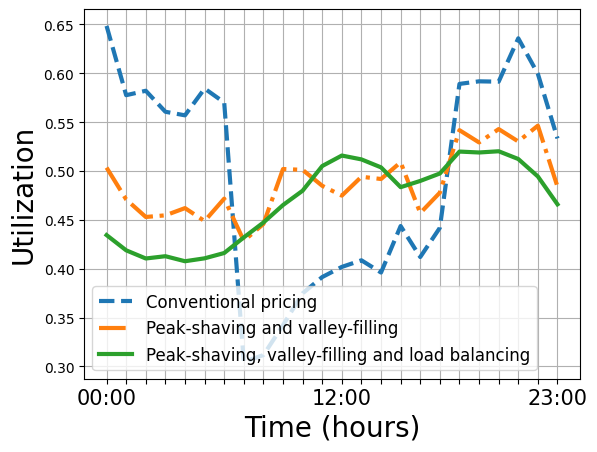}
         \caption{Residential area}
         \label{fig: res_util}
    \end{subfigure}
    \begin{subfigure}{0.24\textwidth}
         \includegraphics[width=\textwidth]{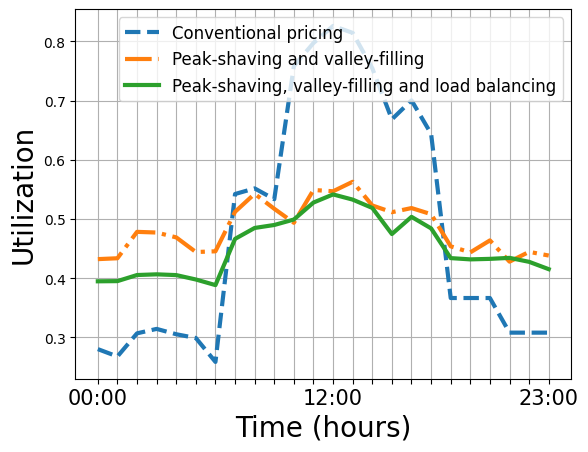}
        \caption{Commercial area}
        \label{fig: com_util}
    \end{subfigure}
    \caption{Effect of Optimized EV charging pricing on utilization of distribution networks}
    \label{fig: com_res_util}
\end{figure}

\subsubsection{Distribution networks utilization and inter-distribution networks load-balancing}

Figures \ref{fig: res_util} and \ref{fig: com_util} show the utilization of the distribution networks in the residential and commercial areas for the fixed maximum capacities of $0.9$ GW and $1.0$ GW respectively. In both areas, the conventional pricing strategy is seen to produce the utilization that varies significantly, with values that swiftly approach the maximum capacity during the peak times, and very low utilization values during the off-peak times.

\begin{figure} [h]
    \centering
    \includegraphics[width=0.5\textwidth]{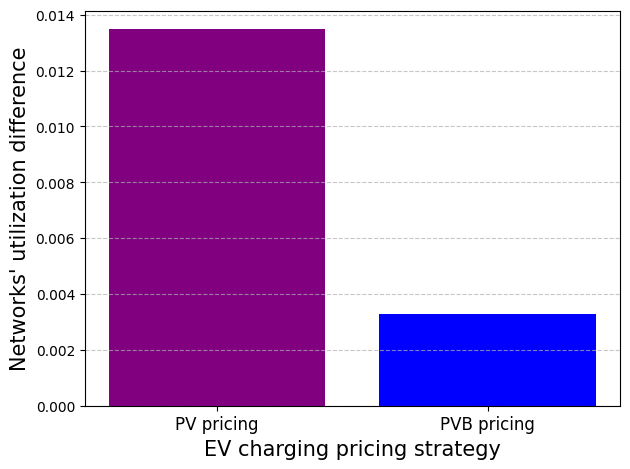}
    \caption{Inter-distribution networks utilization difference under PV and PVB EV charging pricing strategies}
    \label{fig: utils_diff}
\end{figure}

This shows that with the increased penetration of EVs, the distribution networks will reach the maximum capacity and require upgrades for mainly for handling the load at peak times.
On the other hand, the PV EV charging pricing strategy produces better distribution network utilization, which is better improved by the addition of load-balancing component. The load-balancing component adds the load-shifting property that makes the utilization in both networks to approach the uniform value, thus at the given time, a highly utilized distribution network is relieved by the nearby one.


To compare the load-balancing capability of PVB EV charging pricing over PV EV charging strategy, we take the average values of grid utilization for 24-hour period for both the residential and commercial area distribution networks and evaluate the magnitude of the difference. Figure \ref{fig: utils_diff} shows the magnitudes of the differences in utilization of residential and commercial area distribution networks.
Compared to PV, PVB EV charging pricing strategy performs better, with the networks' utilization difference approaching zero. This is particularly very helpful for the neighboring distribution networks that that have the capacities that differ with very high margins.

\section{Conclusion and Future Work}
\label{sec:conclusion}

In this work, electric vehicles are regarded as mobile loads that can be charged in any distribution network that is reasonably close by. Based on the environmental variables, EV charging prices, and the vehicles' travel behaviour, we forecast the conventional load and EV load data in a distribution network. XGBoost outperforms neural networks and polynomial regression for conventional load forecasting, and we use polynomial regression for the EV load forecasting.

Motivated by the fact that EV charging price is a significant determining factor of where and when EVs are charged, we propose a load-balancing strategy between the distribution networks based on dynamic EV charging prices, which are determined by the collaborative DRL agents, with each agent in each network. Three widely used DRL algorithms: DDPG, SAC, and PPO were compared. DDPG outperforms SAC and PPO for this task. Simulation results show that the proposed strategy can handle peak-shaving, valley-filling, and load balancing between distribution networks in close vicinity, thus improving the grid utilization.

This work brings a contribution towards optimal integration of EVs on the smart power grids, especially the grids which were initially not designed to handle high power demands brought by high penetration of EVs into the grid. For future work, more than two distribution networks will be considered, taking into consideration the distance between EVCSs in different distribution networks, both the charging and discharging of the EVs, the integration of renewable energy sources as they are being implemented in practical networks, as well as user compliance risks. On top of the load balancing between the distribution networks, minimization of the power loss and communication latency between the agents will also be considered.

\bibliographystyle{IEEEtran}
\bibliography{ref.bib}

@article{lepolesa2024optimal,
  title={Optimal EV Charging Strategy for Distribution Networks Load Balancing In a Smart Grid Using Dynamic Charging Price},
  author={Lepolesa, Leloko J and Adetunji, Kayode E and Ouahada, Khmaies and Liu, Zhenqing and Cheng, Ling},
  journal={IEEE Access},
  year={2024},
  publisher={IEEE}
}

@article{aghajan2022charging,
  title={Charging and discharging of electric vehicles in power systems: An updated and detailed review of methods, control structures, objectives, and optimization methodologies},
  author={Aghajan-Eshkevari, Saleh and Azad, Sasan and Nazari-Heris, Morteza and Ameli, Mohammad Taghi and Asadi, Somayeh},
  journal={Sustainability},
  volume={14},
  number={4},
  pages={2137},
  year={2022},
  publisher={MDPI}
}

@article{khan2022effects,
  title={Effects of Optimization on User-based Charging/Discharging Control Strategy},
  author={Khan, Zohaib and Wang, Yang},
  journal={Recent Advances in Electrical \& Electronic Engineering (Formerly Recent Patents on Electrical \& Electronic Engineering)},
  volume={15},
  number={2},
  pages={158--170},
  year={2022},
  publisher={Bentham Science Publishers}
}

@article{tariang2023survey,
  title={A Survey on Coordinated Charging Methods for Electric Vehicles},
  author={Tariang, Dathewbhalang and Das, Gitu},
  journal={ADBU Journal of Electrical and Electronics Engineering (AJEEE)},
  volume={5},
  number={1},
  pages={36--48},
  year={2023}
}

@misc{nhts_website ,
author = {U.S. Department of Transportation Federal Highway Administration},
title = {2017 National Household Travel Survey},
year = {2017},
howpublished = {Available at \url{https://nhts.ornl.gov/} (Last accessed: 2025/02/15)}
}

@article{gilleran2021impact,
  title={Impact of electric vehicle charging on the power demand of retail buildings},
  author={Gilleran, Madeline and Bonnema, Eric and Woods, Jason and Mishra, Partha and Doebber, Ian and Hunter, Chad and Mitchell, Matt and Mann, Margaret},
  journal={Advances in Applied Energy},
  volume={4},
  pages={100062},
  year={2021},
  publisher={Elsevier}
}

@article{kim2019insights,
  title={Insights into residential {EV} charging behavior using energy meter data},
  author={Kim, Jae D},
  journal={Energy Policy},
  volume={129},
  pages={610--618},
  year={2019},
  publisher={Elsevier}
}

@inproceedings{sohail2023impact,
  title={Impact Analysis of Time-of-Use pricing enabled Electric Vehicle charging to the uncoordinated charging on a Distribution Network},
  author={Sohail, Muhammad Abeer and Khan, Rashna and Mukhtar, Syed Hammad and Usman, Ahmad},
  booktitle={2023 IEEE Power \& Energy Society General Meeting (PESGM)},
  pages={1--5},
  year={2023},
  organization={IEEE}
}

@inproceedings{nath2024short,
  title={Short-term Electric Vehicle Charging Load forecasting using Transfer and Meta-learning},
  author={Nath, Keshav and Gowda, Shashank Narayana and Zhang, Chen and Gowda, Rohan Shivesh and Gadh, Rajit},
  booktitle={2024 IEEE Power \& Energy Society Innovative Smart Grid Technologies Conference (ISGT)},
  pages={1--5},
  year={2024},
  organization={IEEE}
}

@article{lee2018analysis,
  title={An analysis of price competition in heterogeneous electric vehicle charging stations},
  author={Lee, Woongsup and Schober, Robert and Wong, Vincent WS},
  journal={IEEE Transactions on Smart Grid},
  volume={10},
  number={4},
  pages={3990--4002},
  year={2018},
  publisher={IEEE}
}

@article{kuang2024unravelling,
  title={Unravelling the effect of electricity price on electric vehicle charging behavior: A case study in Shenzhen, China},
  author={Kuang, Haoxuan and Zhang, Xinyu and Qu, Haohao and You, Linlin and Zhu, Rui and Li, Jun},
  journal={Sustainable Cities and Society},
  pages={105836},
  year={2024},
  publisher={Elsevier}
}

@article{kuang2024physics,
  title={A physics-informed graph learning approach for citywide electric vehicle charging demand prediction and pricing},
  author={Kuang, Haoxuan and Qu, Haohao and Deng, Kunxiang and Li, Jun},
  journal={Applied Energy},
  volume={363},
  pages={123059},
  year={2024},
  publisher={Elsevier}
}

@article{qu2024physics,
  title={A physics-informed and attention-based graph learning approach for regional electric vehicle charging demand prediction},
  author={Qu, Haohao and Kuang, Haoxuan and Wang, Qiuxuan and Li, Jun and You, Linlin},
  journal={IEEE Transactions on Intelligent Transportation Systems},
  year={2024},
  publisher={IEEE}
}

@misc{rmse,
author = {Khan Academy},
title = {Standard deviation of residuals or Root-mean-square error {(RMSD)}},
howpublished = {Available at \url{https://www.khanacademy.org/math/statistics-probability/ describing-relationships-quantitative-data/assessing-the-fit -in-least-squares-regression/v/standard-deviation-of-residuals -or-root-mean-square-error-rmsd} (Last accessed: 2024/10/25)}
}

@article{kazemtarghi2024dynamic,
  title={Dynamic pricing strategy for electric vehicle charging stations to distribute the congestion and maximize the revenue},
  author={Kazemtarghi, Abed and Mallik, Ayan and Chen, Yan},
  journal={International Journal of Electrical Power \& Energy Systems},
  volume={158},
  pages={109946},
  year={2024},
  publisher={Elsevier}
}

@article{chakraborty2024planning,
  title={Planning of fast charging infrastructure for electric vehicles in a distribution system and prediction of dynamic price},
  author={Chakraborty, Pratyush and Pal, Mayukha and others},
  journal={International Journal of Electrical Power \& Energy Systems},
  volume={155},
  pages={109502},
  year={2024},
  publisher={Elsevier}
}

@misc{morocco_data,
author = {fedesoriano},
title = {Electric Power Consumption},
howpublished = {Available at \url{https://www.kaggle.com/datasets/fedesoriano/electric-power-consumption} (Last accessed: 2025/02/26)}
}

@misc{minmax ,
author = {Codecademy},
title = {Normalization},
howpublished = {Available at \url{https://www.codecademy.com/articles/normalization} (Last accessed: 2025/02/27)}
}

@article{li2017deep,
  title={Deep reinforcement learning: An overview},
  author={Li, Yuxi},
  journal={arXiv preprint arXiv:1701.07274},
  year={2017}
}

@article{wang2022deep,
  title={Deep reinforcement learning: A survey},
  author={Wang, Xu and Wang, Sen and Liang, Xingxing and Zhao, Dawei and Huang, Jincai and Xu, Xin and Dai, Bin and Miao, Qiguang},
  journal={IEEE Transactions on Neural Networks and Learning Systems},
  volume={35},
  number={4},
  pages={5064--5078},
  year={2022},
  publisher={IEEE}
}

@article{lillicrap2015continuous,
  title={Continuous control with deep reinforcement learning},
  author={Lillicrap, Timothy P and Hunt, Jonathan J and Pritzel, Alexander and Heess, Nicolas and Erez, Tom and Tassa, Yuval and Silver, David and Wierstra, Daan},
  journal={arXiv preprint arXiv:1509.02971},
  year={2015}
}

@inproceedings{chen2016xgboost,
  title={Xgboost: A scalable tree boosting system},
  author={Chen, Tianqi and Guestrin, Carlos},
  booktitle={Proceedings of the 22nd acm sigkdd international conference on knowledge discovery and data mining},
  pages={785--794},
  year={2016}
}

@article{li2024impact,
  title={Impact of electric vehicle charging demand on power distribution grid congestion},
  author={Li, Yanning and Jenn, Alan},
  journal={Proceedings of the National Academy of Sciences},
  volume={121},
  number={18},
  pages={e2317599121},
  year={2024},
  publisher={National Academy of Sciences}
}

@article{nutkani2024impact,
  title={Impact of EV charging on electrical distribution network and mitigating solutions--A review},
  author={Nutkani, Inam and Toole, Hamish and Fernando, Nuwantha and Andrew, Loh Poh Chiang},
  journal={IET Smart Grid},
  volume={7},
  number={5},
  pages={485--502},
  year={2024},
  publisher={Wiley Online Library}
}

@article{abiassaf2024impact,
  title={Impact of EV charging, charging speed, and strategy on the distribution grid: a case study},
  author={Abiassaf, Gerard Albadawi and Arkadan, Abd A},
  journal={IEEE Journal of Emerging and Selected Topics in Industrial Electronics},
  volume={5},
  number={2},
  pages={531--542},
  year={2024},
  publisher={IEEE}
}

@article{strezoski2024enabling,
  title={Enabling mass integration of electric vehicles through distributed energy resource management systems},
  author={Strezoski, Luka and Stefani, Izabela},
  journal={International Journal of Electrical Power \& Energy Systems},
  volume={157},
  pages={109798},
  year={2024},
  publisher={Elsevier}
}

@inproceedings{srividhya2024optimizing,
  title={Optimizing Electric Vehicle Charging Networks Using Clustering Technique},
  author={Srividhya, V and Gowriswari, S and Antony, N Vini and Murugan, S and Anitha, K and Rajmohan, M},
  booktitle={2024 2nd International Conference on Computer, Communication and Control (IC4)},
  pages={1--5},
  year={2024},
  organization={IEEE}
}

@misc{iea_co2,
author = {{International Energy Agency}},
title = {Tracking Transport},
howpublished = {Available at \url{https://www.iea.org/energy-system/transport} (Last accessed: 2025/04/09)}
}

@article{chen2014polynomial,
  title={Polynomial regression models},
  author={Chen, Ray-Bing},
  journal={Lecture Notes of National University of Kaohsiung},
  year={2014}
}

@article{will2016understanding,
  title={Understanding user acceptance factors of electric vehicle smart charging},
  author={Will, Christian and Schuller, Alexander},
  journal={Transportation Research Part C: Emerging Technologies},
  volume={71},
  pages={198--214},
  year={2016},
  publisher={Elsevier}
}

@article{saxena2024power,
  title={Power Quality Enhancement by Mitigating Load Imbalance from Random Electric Vehicle Fleet at Electric Vehicle Charging Stations},
  author={Saxena, Nitin Kumar and Gao, David Wenzhong},
  journal={Green Energy and Intelligent Transportation},
  pages={100222},
  year={2024},
  publisher={Elsevier}
}

@article{das2024advantageous,
  title={An advantageous charging/discharging scheduling of electric vehicles in a PV energy enhanced power distribution grid},
  author={Das, Pritam and Kayal, Partha},
  journal={Green Energy and Intelligent Transportation},
  volume={3},
  number={2},
  pages={100170},
  year={2024},
  publisher={Elsevier}
}

@article{li2024toward,
  title={Toward efficient smart management: A review of modeling and optimization approaches in electric vehicle-transportation network-grid integration},
  author={Li, Mince and Wang, Yujie and Peng, Pei and Chen, Zonghai},
  journal={Green Energy and Intelligent Transportation},
  year={2024}
}

\end{document}